# Towards the Quantized Anomalous Hall effect in AlO$_x$-capped MnBi$_2$Te$_4$

Yongqian Wang[1,2,*], Bohan Fu[1,2,*], Yongchao Wang[3], Zichen Lian[3], Shuai Yang[1,2], Yaoxin Li[3], Liangcai Xu[3], Zhiting Gao[4], Wanjun Jiang[3,5], Jinsong Zhang[3,5,7], Yayu Wang[3,5,6,7], Chang Liu[1,2,†]

[1]*Beijing Key Laboratory of Opto-electronic Functional Materials & Micro-Nano Devices, Department of Physics, Renmin University of China, 100872, Beijing, China*

[2]*Key Laboratory of Quantum State Construction and Manipulation (Ministry of Education), Renmin University of China, Beijing, 100872, China*

[3]*State Key Laboratory of Low Dimensional Quantum Physics, Department of Physics, Tsinghua University, Beijing 100084, China*

[4]*Beijing Academy of Quantum Information Sciences, Beijing 100193, China*

[5]*Frontier Science Center for Quantum Information, Beijing 100084, China*

[6]*New Cornerstone Science Laboratory, Frontier Science Center for Quantum Information, Beijing 100084, P. R. China*

[7]*Hefei National Laboratory, Hefei, 230088, China*

[*] These authors contributed equally to this work.

[†] Emails: liuchang_phy@ruc.edu.cn;

**The quantum anomalous Hall effect in layered antiferromagnet MnBi$_2$Te$_4$ harbors a rich interplay between magnetism and topology, holding a significant promise for low-power electronic devices and topological antiferromagnetic spintronics. In recent years, MnBi$_2$Te$_4$ has garnered considerable attention as the only known material to exhibit the antiferromagnetic quantum anomalous Hall effect. However, this field faces significant challenges as realizing quantized transport at zero magnetic fields depends critically on fabricating high-quality device. In this article, we address the detrimental influences of fabrication on MnBi$_2$Te$_4$ by simply depositing an AlO$_x$ thin layer on the surface prior to fabrications. Optical contrast and magnetotransport measurements on over 50 samples demonstrate that AlO$_x$ can effectively preserve the pristine state of the samples and significantly enhance the anomalous Hall effect towards quantization. Scaling analysis reveals the Berry curvature dominated mechanism of the anomalous Hall effect at various magnetic configurations. By adjusting the gate voltage, we uncover a gate independent antiferromagnetism in MnBi$_2$Te$_4$. Our experiment not only pave the way for fabricating high-quality transport devices but also advance the exploration of exotic quantum physics in 2D materials.**

Magnetic topological materials have emerged as a frontier in condensed matter physics, providing promising platforms for exploring exotic quantum phenomena and potential applications in topological spintronics[1-3]. In experimental studies, fabricating high-quality devices with quantized transport is significant for uncovering novel topological physics. As the first identified material possessing 2D characteristics, intrinsic magnetic order, and band topology simultaneously, MnBi$_2$Te$_4$ not only exhibits rich novel phenomena when exfoliated down to few-layer limit[4-6], but is also considered capable of addressing the disorder issue prevalent in conventional doped topological insulators (TIs)[7]. The bulk of MnBi$_2$Te$_4$ can be regarded as a stacking of Te-Bi-Te-Mn-Te-Bi-Te septuple layer (SL) along the *z*-direction (Fig. 1a). A-type antiferromagnetic (AFM) structure with interlayer AFM order and intralayer ferromagnetic (FM) order forms at the Néel temperature ($T_N$) of ~ 25 K. The layer-sensitive magnetic order gives rise to a rich variety of topological quantum states and exotic magnetoelectric responses while interacted with topology[8-16]. In odd-SL MnBi$_2$Te$_4$, the half-quantized surface Hall conductivity $\sigma_{xy}$ of the same sign drives the system to the quantum anomalous Hall (QAH) state with dissipationless edge state transport[8] (Fig. 1b). This manifests as a quantized $\sigma_{xy} = Ce^2/h$ at zero magnetic fields ($\mu_0 H = 0$), where $C$ is the Chern number, $e$ is the electron charge, and $h$ is Planck constant. In even-SL MnBi$_2$Te$_4$, the opposite surface magnetization results in the axion insulator state characterized with a zero Hall plateau[10]. Recent progresses in MnBi$_2$Te$_4$ have unveiled a plethora of novel topological phenomena, including the Mobius insulator[17], layer Hall effect[11], axion optical induction[18], and quantum metric nonlinear transport[19,20].

Despite the observation of the QAH effect and axion insulator in 5- and 6-SL MnBi$_2$Te$_4$, the $T$ at which quantization is achieved remains much lower than its $T_N$. A more formidable challenge arises from the exceptionally low yield of fabricating MnBi$_2$Te$_4$ devices that exhibit quantized transport[8]. In the past few years, neither the perfect quantized nor zero Hall plateau has been easily reproduced. The lack of quantization not only obstructs the discovery of new physical phenomena but also complicates the interpretation of data. Possible reasons include various defect structures and impurity phases[21-25], instability of surface electronic structure[26-29], and weakened surface anisotropy[29-31] in bulk crystals. A recent study that combines optical contrast ($O_c$), transport, and magneto-optical Kerr effect measurements revealed a substantial influence of fabrication on the properties of MnBi$_2$Te$_4$ devices[32]. The contact with photoresist not only reduces $O_c$ during the fabrication, but also may leads to mismatched even-odd-layer dependent magnetotransport[32]. Developing a low-damage fabrication method has emerged as an urgent and practical imperative in the fields of both topological quantum matter and 2D materials.

In this work, we optimize the conventional fabrication process by thermally depositing an AlO$_x$ protective layer on MnBi$_2$Te$_4$ surface prior to the standard electron beam lithography (EBL). Through $O_c$ measurements on a series of MnBi$_2$Te$_4$, we find that this issue caused by Polymethyl Methacrylate (PMMA) photoresist is largely mitigated. Transport measurements in 17 devices suggests that the AH effects in AlO$_x$ capped devices are substantially enhanced towards $e^2/h$. Further $T$ and gate voltages ($V_g$s) dependent experiments demonstrate the Berry curvature dominated mechanism of the AH effect at different magnetic configurations. The variation of critical exponent ($\beta$), $T_N$, and coercive field ($H_c$) as $V_g$ reveals a carrier independent antiferromagnetism in MnBi$_2$Te$_4$. Our experiment proposes a simple low-damage scheme for the fabrication of high-quality topological transport devices, paving the way for realizing the QAH effect and the exploration of novel topological quantum phenomena.

**Results**

**Device fabrication and optical contrast**

We got inspirations from previous experiments in which those MnBi$_2$Te$_4$ devices exhibiting pronounced AH effect usually had Al$_2$O$_3$ on the bottom of the flake incorporated in their devices[8,33,34]. This similarity indicates that the contact with Al$_2$O$_3$ may aids in improving the quality of MnBi$_2$Te$_4$ device. Combined with our recent finding of the effect of fabrication on the top surface of MnBi$_2$Te$_4$, we come up with a simple yet effective idea that by depositing an AlO$_x$ layer on top of MnBi$_2$Te$_4$ to achieve the QAH effect. Figure 1c shows the schematic of the fabrication process. First, we transferred thick MnBi$_2$Te$_4$ from a bulk crystal onto the substrates using a scotch tape. We then employed the mechanical exfoliation method to obtain the flakes with target thicknesses. The one-to-one correspondence between $O_c$ and thickness enables the rapid determination of layer number by optical method[11]. Subsequently, a 3-nm AlO$_x$ layer was deposited by thermal evaporation. We then used the standard EBL to expose the desired Hall bar. Next, the AlO$_x$ above the designed electrode regimes was etched away using Ar ion etching, followed by the deposition of Cr/Au (3/50 nm) electrodes. Finally, a layer of PMMA was coated for further protection. The details of the fabrication are described in the Methods section. Compared to the Al$_2$O$_3$-assisted exfoliation and stencil mask method[8], our method is based on the standard EBL process, which allows for the fabrication of specific fine structures. Moreover, because the AlO$_x$ layer is on the top surface, it offers an effective protection while also enables future extension as a top gate.

Figure 1d shows a schematic of a Hall bar device covered with AlO$_x$ capping layer and its cross-sectional view. To investigate the influence of AlO$_x$ on MnBi$_2$Te$_4$, we first compare the optical properties of $MnBi_2Te_4$ flakes with varied thicknesses, which were exfoliated from the same single crystal (see supplementary Fig. S1 for the $O_c$ change in each step). The optical images in Fig. 1e clearly show that for samples without $AlO_x$ (up panel), the color of all four regions changes significantly before and after contact with PMMA. In contrast, those regimes with $AlO_x$ (down panel) do not exhibit noticeable change during the same process. To further study the effect of $AlO_x$ quantitatively, we extract the $O_c$ values and compare their variations directly. As plotted in Fig. 1f, a significant reduction of $O_c$ in all four regimes without $AlO_x$ are observed. In contrast, $O_c$ remains nearly unchanged for the four regimes with $AlO_x$ layer. To further determine the universal role of $AlO_x$, we measured the $O_c$ of 47 samples exfoliated from the same crystal. As displayed in Fig.1g, the variation in $O_c$ falls well into two regions (red and blue), representing unchanged thickness and a reduction of 1 SL, respectively. These behaviors demonstrate that $AlO_x$ capping layer can effectively mitigates the damage caused by PMMA.

**Statistical Analysis of the Influence of Fabrication Method on Transport Behaviors**

In magnetic topological systems, the AH effect typically results from three mechanisms: intrinsic Berry curvature $\Omega(\boldsymbol{k})$, skew-scattering, and side-jump. In the transport of $MnBi_2Te_4$, due to defects or impurity phases, all three mechanisms could contribute to the AH effect[35,36]. However, in an ideal quantized Hall system, the transverse transport should be dominated by $\Omega(\boldsymbol{k})$ in $\boldsymbol{k}$ momentum space[37]. Theoretically, $\sigma_{xy}$ can be calculated by integrating $\Omega$ over $\boldsymbol{k}$, as expressed by:

$$\sigma_{xy} = -\frac{e^2}{2\pi h}\int \Omega(\boldsymbol{k})d^2\boldsymbol{k}$$

When the Fermi level ($E_F$) is tuned into the magnetic exchange gap, the integral of $\Omega$ equals the $C$ number multiplied by $2\pi$, resulting in the quantization of $\sigma_{xy}$ at $e^2/h$. To investigate the influence of $AlO_x$ on the intrinsic AH effect, we measured the transport behaviors of 17 odd-SL $MnBi_2Te_4$ devices. All the data present in the main text was obtained at the charge neutral point (CNP) unless otherwise specified. Prior to this, we measured the current-voltage curve of the $AlO_x$ layer to exclude its contribution to transport (see supplementary Fig. S2). Figures 2a-2d show the $\mu_0H$ dependence of $\sigma_{xy}$ and $\sigma_{xx}$ for two 7-SL devices exfoliated from the same thick flake on the same tape. Both samples exhibit quantized $\sigma_{xy}$ at high $\mu_0H$ when entering the Chern insulator state ($C$ = -1). However, their AH effect at zero field exhibits dramatically different behaviors. For the sample without AlO$_x$ (Fig. 2a), $\sigma_{xy}$ almost vanishes at zero fields, reminiscent of the zero Hall plateau in an even-SL axion insulator. Such result is consistent with our previous finding that fabrication may damage the top surface, leading to a reduction of effective thickness by one SL[32]. However, for the device fabricated by our new method, a large $\sigma_{xy}$ with well squared hysteresis is observed. The insets illustrate the possible evolution of the topological surface state (green) without and with AlO$_x$. Figures 2b and 2d summarize the $\sigma_{xy}$ and $\sigma_{xx}$ at $\mu_0H$ = -8 T and 0 for the two samples as a function of $V_g$. For device #1, $\sigma_{xy}$ exhibits an accidental zero Hall point during sweeping $V_g$. In sharp contrast, for device #10, $\sigma_{xy}$ shows a wide plateau in the $V_g$ range of Chern insulator, indicating an incipient QAH state.

In 2D materials, the transport behaviors of thin flakes are inevitably affected by sample quality fluctuations. Previous studies have shown that MnBi$_2$Te$_4$ exhibits sample-dependent properties, even for devices prepared from the same crystal[8,22,23]. To further demonstrate the increase of the AH effect, we compare the transport properties of 17 MnBi$_2$Te$_4$, as shown in Figs. 2e and 2f. The samples were numbered based on the increasing order of their $\sigma_{xy}$ at $\mu_0H$ = 0. Apart from devices #2, #4, #6, and #8, all the other 13 devices were cleaved from crystal 2. Nevertheless, the 4 devices represent the samples with the largest AH effect that have been exfoliated from crystal 1. All the samples enter the Chern insulator and exhibit $\sigma_{xy} = e^2/h$ at high $\mu_0H$, indicating the overall high quality of our crystals. Interestingly, all 9 samples without AlO$_x$ exhibit small $\sigma_{xy}$ and indiscernible hysteresis. In contrast, the other 8 samples with AlO$_x$ display large $\sigma_{xy}$ and square-shaped hysteresis, with two samples (#16 and #17) showing the QAH effect. Figure 2g summarizes the $\sigma_{xy}$ at $\mu_0H$ = 0 of these samples, which clearly suggests that AlO$_x$ plays a crucial role in optimizing the QAH effect. Remarkably, even without full quantization, the AH effect in the samples with AlO$_x$ has already surpassed the values reported for most odd-SL MnBi$_2$Te$_4$ in current literatures[9,11-14,16,32,35].

The fabrication of high-quality devices enables us to compare the influences of magnetic properties on transport. Figures 3a to 3c show the $\mu_0H$ dependent $\sigma_{xx}$ and $\sigma_{xy}$ for three samples obtained from the same crystal. Fortunately, for devices #11 and #16, they were obtained on the same substrate during one cleaving process. It enables us to further explore the influence of $AlO_x$ on $MnBi_2Te_4$ while preserving the consistency of the sample as much as possible. In device #11, an additional 30 nm $AlO_x$ was deposited after the first deposition along with #16. Overall, the three devices display similar quantized behavior, with the main differences being the values of $\sigma_{xx}$ and $\sigma_{xy}$ at $\mu_0 H = 0$. However, the sharpness of the plateau transition, which reflects the magnetic flipping process, differs dramatically. For device #11 with a longer time of $AlO_x$ deposition, the $\sigma_{xy}$ and $\sigma_{xx}$ at $\mu_0 H = 0$ are 0.5 and 1.1 $e^2/h$, respectively, and the plateau transition appears with a relatively gentle process. Whereas for device #16, though the value of $\sigma_{xx}$ does not change, $\sigma_{xy}$ is much improved and approaches $e^2/h$. The plateau transition is also sharper. Device #17 fully enters the QAH state, with $\sigma_{xy}$ reaching $e^2/h$ and $\sigma_{xx}$ dropping to zero. The sharper plateau transition along with better QAH state indicate there may be a better perpendicular magnetic order in device #17.

The scaling relation between $\sigma_{xy}$ and $\sigma_{xx}$ may further help us understand the role of $AlO_x$ in enhancing QAH effect. Figures 3d to 3f show the variation of $\sigma_{xy}$ as a function of $\sigma_{xx}$ during the cooling process under different $\mu_0 H$ and $V_g$s. As the AFM order strengthens at low $T$s, $\sigma_{xy}$ begins to exhibit behavior independent of $\sigma_{xx}$ and gradually approaches quantization, which is of typical the scaling behavior for $\Omega(\boldsymbol{k})$ dominated mechanism[37]. Upon increasing $\mu_0 H$, the sample undergoes AFM, canted AFM, and finally enters the FM state, accompanied with $\sigma_{xy}$ saturating at $e^2/h$ at higher $T$s. For device #11 with relatively weaker out-of-plane order, the exchange gap is smaller, thermal fluctuations can more easily smear out the role of $\Omega(\boldsymbol{k})$ (top in Fig. 3g). Strict quantization appears only when all moments are parallelly aligned because the gap is overall positively correlated with magnetization[38]. In contrast, for the sample with better magnetic order, such as device #17, the larger gap allows for $\sigma_{xy}$ quantization even in the AFM state despite a small net moment. The effect of magnetic configuration on exchange gap and $\Omega(\boldsymbol{k})$ is illustrated in Fig. 3g. The red and blue represent the distribution of the $\Omega(\boldsymbol{k})$ in the conduction and valence bands, respectively. The $\Omega(\boldsymbol{k})$ in larger gapped system exhibits greater robustness against thermal fluctuations.

**Gate voltage independent Magnetism**

Next, we investigate the effect of $V_g$ on the magnetic properties of MnBi$_2$Te$_4$. Previous studies on magnetically doped TIs have revealed different $V_g$ dependent behaviors for critical $T$ and $H_c$ (ref. [39-41]). The electrical control of 2D magnetism in layered magnets also attracted significant interests[42,43]. As the first layered topological antiferromagnet, it remains yet to be determined whether $V_g$ can exert similar effects in MnBi$_2$Te$_4$. Figures 4a and 4b display the $\mu_0H$ dependent $\sigma_{xy}$ and $\sigma_{xx}$ for device #16 at varied $T$s. The Hall quantization and hysteresis vanishes at around $T = 21$ K, accompanied by the disappearance of $\sigma_{xx}$ peaks at the magnetic phase transition (indicated by triangles). To quantitatively investigate the changes in the AFM state, we extract $H_c$ at different $V_g$s and plot them as a function of $T$ (Fig. 4c). Following the strategy employed in the studies of CrI$_3$, we perform a similar fitting to the critical behaviors using $(1-T/T_N)^\beta$ (ref. [42]). We find that $V_g$ has almost negligible effect on the AFM order. Figure 4d shows the $V_g$ dependent $\beta$ and $T_N$. $T_N$ remains a constant at ~ 21.3 K while $\beta$ consistently maintains around 0.52. Similar results of $\beta$ were observed in previous neutron diffraction and reflectance magneto-circular dichroism studies for bulk[44] and thick flakes[15]. Our experiments further uncover this critical behavior cannot be tuned by single $V_g$. Figure 4e shows a colormap of $\sigma_{xy}$ as a function of $V_g$ and $\mu_0H$. It clearly suggests that $H_c$ remains a constant with $V_g$. Reproducible results from device #11 are shown in supplementary Figs. S3 and S4.

**Discussion**

Finally, we discuss the possible mechanisms underlying the enhancement of AH effect. In our previous research, we found that the coating of PMMA during the EBL process reduces the $O_c$ of MnBi$_2$Te$_4$, leading to a reduction of effective thickness[32]. AlO$_x$ isolates the surface from contact with the resist, thus providing a protection for the sample. However, AlO$_x$ is not the only choice for protection, for example, $h$-BN is more commonly adopted in 2D materials studies. A shadow mask can also be employed to avoid the contact with PMMA. Nevertheless, many experiments have shown that even with these methods[13,19,33,35], the AH effect remains nonquantized. Therefore, AlO$_x$ must play a more critical role in realizing the QAH effect.

Our scaling relation studies imply that the enhancement of perpendicular magnetic order may be crucial for realizing the QAH effect. Based on the above results, we discuss the likely mechanism by which AlO$_x$ improves the magnetism. One conceivable scenario is the electric field strengthened magnetism at the AlO$_x$/MnBi$_2$Te$_4$ interface. As AlO$_x$ alters the inversion symmetry of the surface, an electric field could be created through spin-orbit coupling, which is also a widely adopted method to tune magnetism[45,46]. However, this scenario is inconsistent with our $V_g$ dependent experiment, in which $V_g$ has a relatively small influence on magnetism. Therefore, even if an electric field exists, its contribution in the present study is minor.

Another possible scenario is the strengthened perpendicular magnetic anisotropy (PMA) by AlO$_x$. Theoretical calculation has shown that the MA in monolayer MnBi$_2$Te$_4$ is weak due to the weak $p$-$d$ hybridization between Mn and Te[47]. Inelastic neutron scattering further shows that interlayer two-ion anisotropy can greatly enhances the MA in MnBi$_2$Te$_4$ bulk[30]. However, due to the absence of a neighbor, the top surface MA is much reduced. Therefore, it is likely that AlO$_x$ increases the PMA of AlO$_x$/MnBi$_2$Te$_4$ interface. Remarkably, in spintronics, many experiments have demonstrated that depositing AlO$_x$ on the surface can substantially increase the interfacial PMA[48-50]. The physics arises from the hybridization effect between $d$ orbitals of magnetic ion and $s$-$p$ orbitals of the oxide. Interestingly, it was found that the influence of AlO$_x$ on PMA weakens when the oxidation layer is too thick or the oxidation time is too long. This also consists with our observation that the device with long AlO$_x$ deposition time shows a worse quantization. Notably, although this scenario of enhanced PMA can well explain our experiments, the mechanism by which AlO$_x$ enhances the QAH effect remains unclear so far. A recent theoretical calculation has proposed that covering MnBi$_2$Te$_4$ with a polar insulator can modify the surface potential, thus helps achieving the QAH effect[28]. As a polar insulator, Al$_2$O$_3$ may play an alternative role in realizing the quantization. More studies are needed to figure out this exactly mechanism.

In summary, we report the realization of the QAH effect in AlO$_x$-capped MnBi$_2$Te$_4$. We propose a new fabrication approach based on the standard EBL process. By simply depositing an AlO$_x$ layer on top of MnBi$_2$Te$_4$ surface, there is a substantial enhancement in the AH effect, ultimately reaching quantization. Our experiment addresses a long-standing issue in the field of MnBi$_2$Te$_4$, paving the way for the fabrication of high-quality devices and the investigation of the interplay between topology and layered magnetism. This simple revision in fabrication is of great significance for exploring the intrinsic properties of MnBi$_2$Te$_4$ and would advance its potential applications in topological spintronics.

**Methods**

**Crystal growth** High-quality MnBi$_2$Te$_4$ single crystals were synthesized by directly mixing Bi$_2$Te$_3$ and MnTe with the ratio of 1:1 in a vacuum-sealed silica ampoule. For crystal #1, the mixture was first heated up to 700 °C, and then slowly cooled down to 591 °C, followed by a long period of annealing process. The phase and crystal structure were examined by X-ray diffraction on a PANalytical Empyrean diffractometer with Cu Kα radiation. For crystal #2, a small amount of Te was added to the mixture, with the ratio of Bi$_2$Te$_3$, MnTe and Te modified to 1:1:0.2. The ampoule was slowly heated to 900°C and maintained at this temperature for 1 hour. Subsequently, it was cooled down to 700°C, holding for 1 hour and then gradually cooled to 585°C and maintained for 12 days. After the annealing, the ampoule was quenched in water to avoid phase impurities. Apart from devices #2, #4, #6, and #8, all other 13 devices were cleaved from crystal #2, and these 4 devices also exhibit the largest AH conductivity in the samples prepared from crystal #1.

**Device fabrication** MnBi$_2$Te$_4$ flakes were mechanically exfoliated onto 285 nm thick SiO$_2$/Si substrates by using the scotch tape method in an Ar-filled glove box with O$_2$ and H$_2$O levels lower than 0.1 ppm. Initially, the substrate was thoroughly cleaned with acetone, isopropanol, and deionized water. Then the surface of SiO$_2$/Si was treated with air plasma at approximately 125 Pa for three minutes. The tape-covered substrate was heated up to 60°C for three minutes to facilitate smooth exfoliation of the single crystals into flakes. Micrometer-sized thin flakes can be obtained by mechanically exfoliation on thick flakes for several times. The thickness was identified by optical contrast measurement in the glovebox immediately after exfoliation. After the target flakes were obtained, a nominal 3 nm thick layer of aluminum was deposited onto the surface using a thermal evaporator with a deposition rate of 0.4 A/s under a vacuum better than $4\times10^{-4}$ Pa. Oxygen was then introduced into the chamber, and the aluminum layer was oxidized for five minutes at a pressure of $2\times10^{-2}$ Pa. For device #11, an extra deposition process with longer time was employed to compare the influence of different AlO$_x$ parameter on the transport.

To assess the effect of PMMA on the MnBi$_2$Te$_4$ samples, 270 nm thick PMMA was spin-coated onto the samples in an Ar-filled glovebox at a controlled speed of 4000 round/minute. The samples were then heated at 60°C for seven minutes and left to stabilize in the glove box for 24 hours. Subsequently, the samples were than immersed in acetone for 20 minutes, rinsed with acetone followed by isopropanol, and their optical contrasts were measured immediately after the removal of PMMA. Standard EBL was employed on MnBi$_2$Te$_4$ samples to pattern the Hall bar structure. The oxidized aluminum was first etched from the sample surface using an Ar ion milling machine at a pressure of $2\times10^{-4}$ Torr for 75 seconds. Cr/Au electrodes (3/50 nm) were then deposited using a thermal evaporator connected to a glovebox. Following this, the samples were again spin-coated with PMMA adopting the same parameters as before for further protection.

**Transport measurement** Standard four probe transport measurements for devices #1 to #16 were carried out in a cryostat with the lowest $T$ of ~ 1.5 K and an out-of-plane magnetic field up to ~ 9 T. The longitudinal and Hall voltages were acquired simultaneously via two lock-in amplifiers with an AC current (200 nA, 13 Hz) generated by a Keithley 6221 current source meter. For device #17, the transport was performed in a dilution refrigerator with AC current excitation of 10 nA at 13 Hz. To correct the geometrical misalignment, both the longitudinal and Hall signals were symmetrized and antisymmetrized with respect to magnetic field. The back-gate voltage was applied by a Keithley 2400 source meter through the Si/SiO$_2$ substrate.

**Data Availability:** All data supporting the finding in the study are presented within the main text and the supplementary information. All data are available upon reasonable request from the corresponding author.

**Acknowledgement:** We appreciate the high-quality crystals provided by Yang Wu during the project. Chang Liu was supported by fundings from National Natural Science Foundation of China Grant No. 12274453 and Open Research Fund Program of the State Key Laboratory of Low-Dimensional Quantum Physics Grant No. KF202204. Jinsong Zhang was supported by National Natural Science Foundation of China (Grants No. 12274252 and No. 12350404). Yayu Wang was supported the Basic Science Center Project of Natural Science Foundation of China Grant No. 52388201, the New Cornerstone Science Foundation through the New Cornerstone Investigator Program and the XPLORER PRIZE. Yayu Wang, Jinsong Zhang, and Chang Liu acknowledge the financial support from Innovation program for Quantum Science and Technology Grant No. 2021ZD0302502.

**Author contributions:** C. L. conceived the project. C. L., Y. Y. W., J. S. Z., and W. J. J. supervised the research. Y. C. W grew the MnBi$_2$Te$_4$ crystals, Y. Q. W., B. H. F., and Z. C. L. fabricated the devices and performed the transport measurements with the help of Y. C. W., S. Y., Y. X. L., L. C. X. and Z. T. G., C. L. and Y. Q. W. prepared the manuscript with comment from all authors.

**Competing interests:** The authors declare no competing interests.

# Figure Captions

**Fig. 1 | Fabrication and optical contrast characterization of few-layer MnBi$_2$Te$_4$ flakes.** **a**, Crystal structure of MnBi$_2$Te$_4$. **b**, Schematic of the QAH effect in an odd-SL MnBi$_2$Te$_4$. **c**, Illustration of the device fabrication process. The method is developed based on the standard EBL process. By simply depositing a thin layer of AlO$_x$ on the MnBi$_2$Te$_4$ surface, the PMMA resist is isolated from the top surface. The high insulation and compactness of AlO$_x$ make it possible to fabricate Hall bar patterns while protecting the sample from chemical reagents. **d**, Front and side views of transport device. **e**, Optical images of MnBi$_2$Te$_4$ thin flakes exfoliated from the same crystal. The top (bottom) panel compares the color change of MnBi$_2$Te$_4$ flakes without (with) AlO$_x$ capping before and after contact with PMMA, respectively. **f**, Variation of $O_c$ in different areas. $O_c$ is defined as $(I_{flake} - I_{substrate})/I_{substrate}$, where $I_{flake}$ and $I_{substrate}$ are the intensity of MnBi$_2$Te$_4$ and substrate, respectively. **g**, Statistical analysis of $O_c$ in 47 MnBi$_2$Te$_4$ flakes with (red) and without (blue) AlO$_x$ capping layer. The red and blue lines represent the $O_c$ reduction by 0 and 20 %, respectively, corresponding to no variation in the effective thickness and a decrease by one SL, respectively.

**Fig. 2 | Comparison of transport behaviors of devices obtained by different preparation methods. a**, Transport behavior at CNP for a 7-SL device without AlO$_x$ covering layer. Due to fabrication effects, the surface state shifts down to the second SL (inset), and the hysteresis of $\sigma_{xy}$ near zero field almost disappears. **b,** Variation of $\sigma_{xy}$ and $\sigma_{xx}$ with $V_g$ under $\mu_0 H$ = -8 T and 0, respectively. **c,** Transport behavior of a 7-SL sample exfoliated from a MnBi$_2$Te$_4$ flake on the same scotch tape, but with an AlO$_x$ layer deposited during the fabrication process. The large hysteresis indicates excellent protection of device performance. The inset illustrates that under the protection of AlO$_x$, the topological surface states remain predominantly distributed on the outermost surface. **d,** In the same $V_g$ range of high field Chern insulator state, the $\sigma_{xy}$ at $\mu_0 H$ = 0 exhibits a broad plateau during sweeping $V_g$. **e-f,** $\mu_0 H$ dependent $\sigma_{xy}$ at $T$ = 1.5 K for 17 odd-SL MnBi$_2$Te$_4$ devices at their CNPs. The only difference in their fabrication lies in whether the surface was deposited with AlO$_x$. All the samples exhibit quantized $\sigma_{xy}$ at high

$\mu_0 H$, as indicated by the blue dashed lines. **g,** Summary of the zero field $\sigma_{xy}$ for the 17 devices. Devices with $AlO_x$ capping layer generally exhibit a larger AH effect than those without $AlO_x$.

**Fig. 3 | Scaling relation between $\sigma_{xy}$ and $\sigma_{xx}$ of the intrinsic AH effect. a-c,** $\mu_0 H$ dependent $\sigma_{xy}$ and $\sigma_{xx}$ for three 7-SL devices exfoliated from the same crystal. Devices #11 and #16 were obtained simultaneously in one cleaving process on the same substrate, with the former one undergoing a longer $AlO_x$ deposition time, having a nominal thickness of 33 nm. Device #17 fully enters the QAH state, with $\sigma_{xy}$ quantized in $e^2/h$ and $\sigma_{xx}$ dropping to zero. **d-f,** Evolution of $\sigma_{xy}$ with $\sigma_{xx}$ during the cooling process. As AFM order forms with lowering $T$s, the scaling between $\sigma_{xy}$ and $\sigma_{xx}$ at different $V_g$s gradually collapse into a single curve, and $\sigma_{xy}$ approaches $e^2/h$. The $\sigma_{xx}$ independent behavior reflects the typical Berry curvature-dominated mechanism of the AH effect. **g,** Schematic of the distribution of Berry curvature. From top to bottom, as the AFM order is tuned to the FM order, the exchange gap increases, and the Berry curvature exhibits greater robustness against thermal fluctuations.

**Fig. 4 | Transport and magnetic properties tuned by $V_g$. a-b,** $\mu_0 H$ dependent $\sigma_{xy}$ and $\sigma_{xx}$ at the CNP for device #16 at various $T$s. The hysteresis and the double peak in $\sigma_{xx}$ disappear at around $T = 21$ K. The black triangles mark the position of $H_c$ at different $T$s. **c,** $H_c$ extracted from the field sweep data as a function of $T$ at varied $V_g$s. The solid squares are the data points. The black lines are the data fittings in the form of $(1-T/T_N)^\beta$. **d,** Summarized fitting results $T_N$ and $\beta$ as functions of $V_g$. $T_N$ and $\beta$ are found to be around 21.2 K and 0.52, respectively, which are independent of $V_g$. **e,** Colormap of $\sigma_{xy}$ in the parameter space of $\mu_0 H$ and $V_g$. The boundary between blue and orange region marks the $V_g$ independent $H_c$.

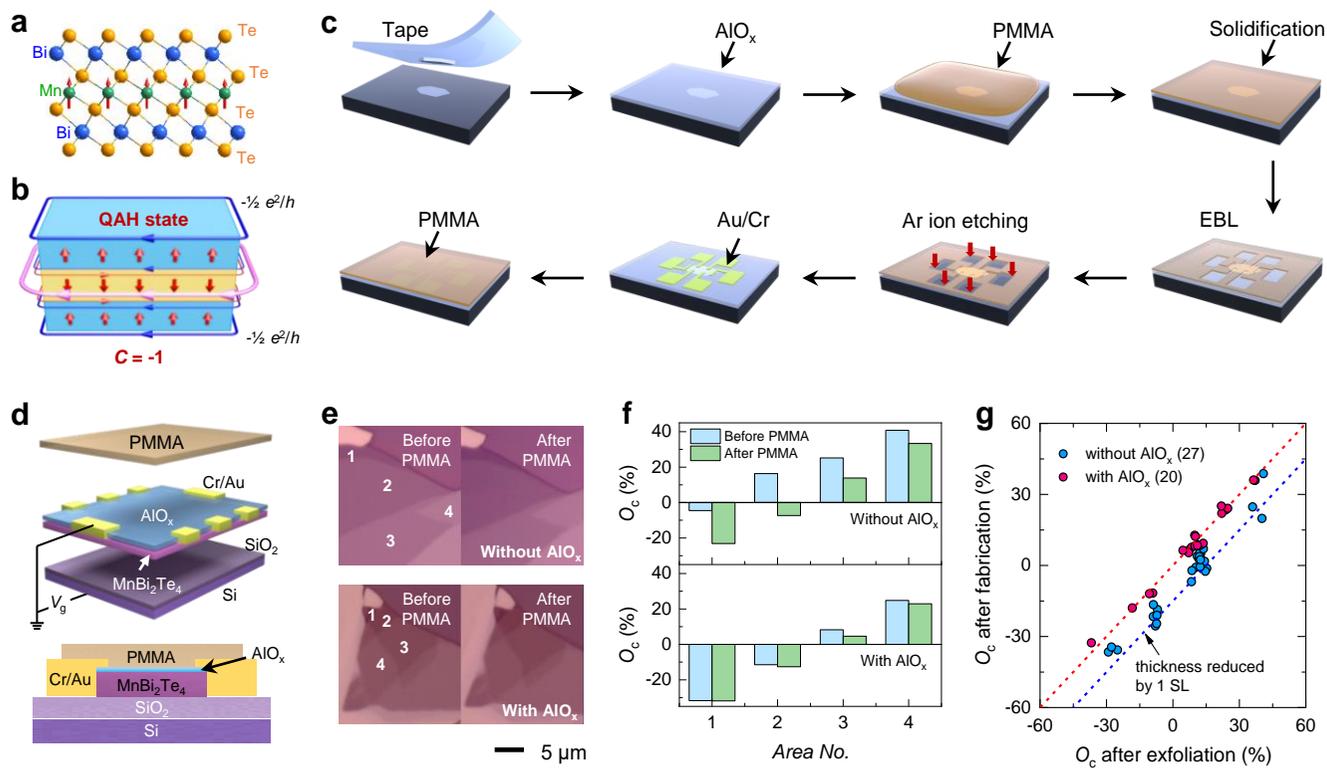

**Figure 1**

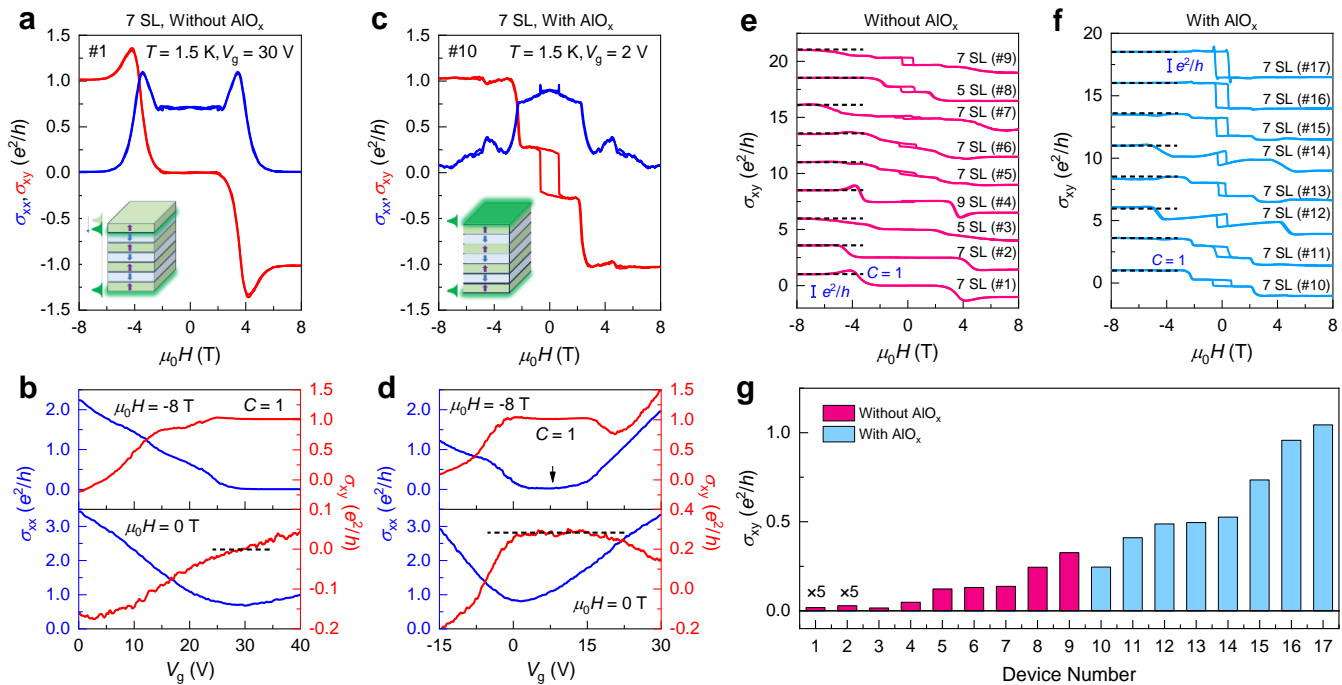

**Figure 2**

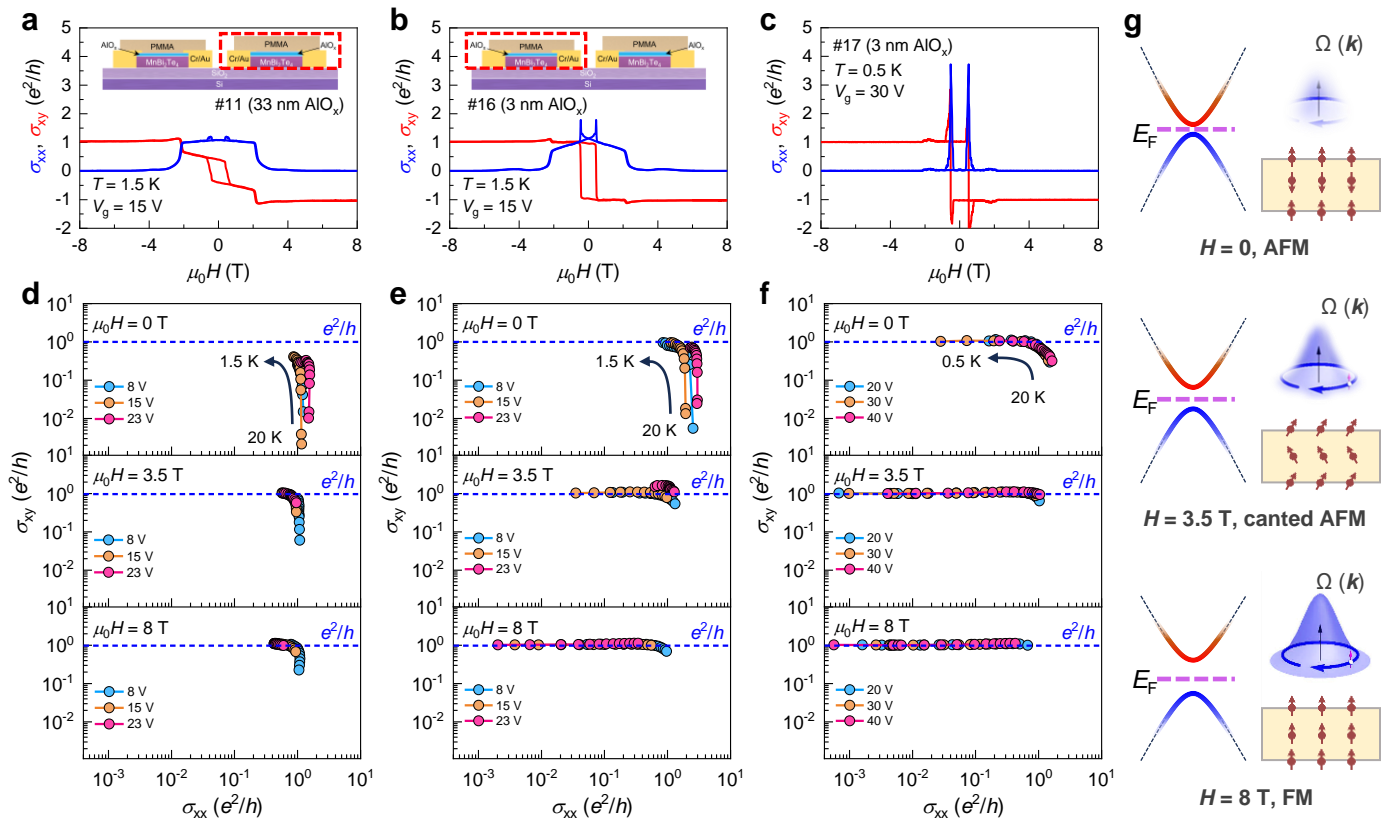

Figure 3

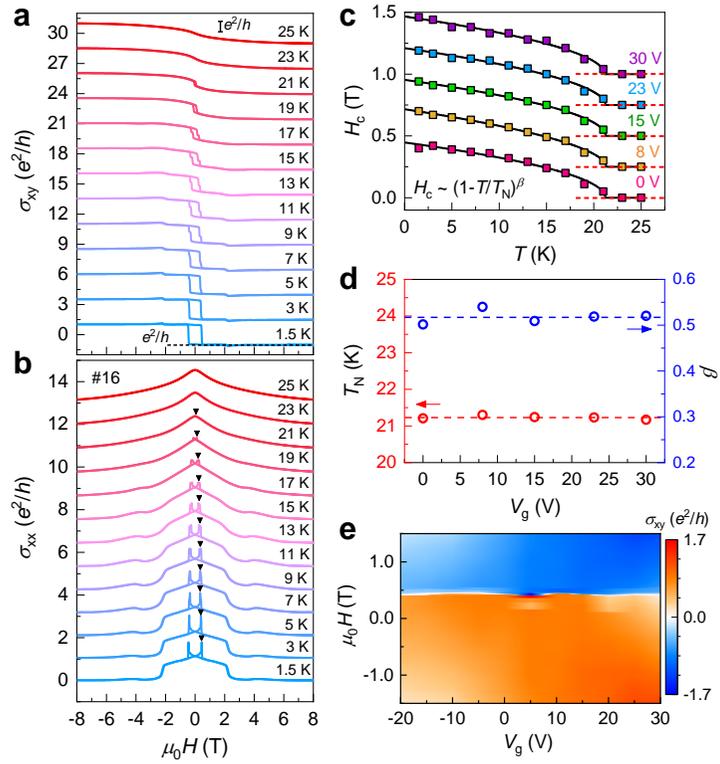

**Figure 4**